\documentclass[a4paper,aps,nofootinbib,onecolumn]{revtex4}
\RequirePackage[english]{babel}
\RequirePackage[latin1]{inputenc}
\RequirePackage[T1]{fontenc}
\RequirePackage{mathrsfs}
\RequirePackage{amsmath}
\RequirePackage{amssymb}
\RequirePackage{amsbsy}
\RequirePackage{bm}
\begin{document}
%%%%%%%%%%%%%%%%%%%%%%%%%%%%%%%%%%%%%%%%%%%%%%%%%%%%%%%%%%%%%%%%%%%%%%%%%%%%%%%%%%%%%%%%%%%%%%%%%%%
\title{\bf{ELKO and Dirac Spinors seen from Torsion}}
\author{Luca Fabbri\footnote{E-mail: fabbri@diptem.unige.it} 
and Stefano Vignolo\footnote{E-mail: vignolo@diptem.unige.it}}
\affiliation{DIME Sezione Metodi e Modelli Matematici, Universit\`{a} di Genova,\\
Piazzale Kennedy Pad.D, 16129 Genova, ITALY}
\date{\today}
%%%%%%%%%%%%%%%%%%%%%%%%%%%%%%%%%%%%%%%%%%%%%%%%%%%%%%%%%%%%%%%%%%%%%%%%%%%%%%%%%%%%%%%%%%%%%%%%%%%
\begin{abstract}
In this paper, the recently-introduced ELKO and the well-known Dirac spinor fields will be compared; however, instead of comparing them under the point of view of their algebraic properties or their dynamical features, we will proceed by investigating the analogies and similarities in terms of their geometrical character viewed from the perspective of torsion. The paper will be concluded by sketching some consequences for the application to cosmology and particle physics.
\end{abstract}
%%%%%%%%%%%%%%%%%%%%%%%%%%%%%%%%%%%%%%%%%%%%%%%%%%%%%%%%%%%%%%%%%%%%%%%%%%%%%%%%%%%%%%%%%%%%%%%%%%%
\maketitle
%%%%%%%%%%%%%%%%%%%%%%%%%%%%%%%%%%%%%%%%%%%%%%%%%%%%%%%%%%%%%%%%%%%%%%%%%%%%%%%%%%%%%%%%%%%%%%%%%%%
\section*{Introduction}
It is now nearly eighty years that the Dirac spinor field has been defined, and its role in describing fermionic fields is well established and widely accepted; nevertheless, the Dirac spinor field's theoretical elegance and phenomenological adequacy in accounting for the properties of leptons and quarks is no reason to stop people investigating for alternative forms of matter, and relatively recently another form of spinorial field, named ELKO, has been introduced \cite{a-g/1}.

In comparison to the Dirac spinor field, there are two main reasons for this ELKO field to be defined: one reason is theoretical, and it climbs back to the very definition of $\frac{1}{2}$-spin spinor fields, the other is phenomenological, naturally arising from the dynamical properties of ELKOs. The theoretical reason is due to the fact that when we wish to define in the most general way $\frac{1}{2}$-spin spinor fields, we have to face the circumstance that not one but two spinor fields are possible according to the two-fold helicity of the field: having right-handed as well as left-handed semi-spinor fields, and being in a situation in which both must be accommodated into a single spinor field, the most natural way in which this can be done is by requiring that the whole spinor field be an eigenstate of parity, and this would straightforwardly lead to the definition of the Dirac field as it is well known; it is perhaps less known but still quite reasonable to see that this is not the only way, and another way, somewhat complementary, is to require that the whole spinor field be an eigenstate of charge conjugation, in German \textit{Eigenspinoren des LadungsKonjugationsOperators} leading to the acronym ELKO \cite{a-g/2}. Such an algebraic rearrangement of the semi-spinorial components has dynamical consequences in the fact that ELKOs can no longer be solution of a Dirac type of equation, although they are still solution of a Klein--Gordon type of equation \cite{a-l-s/1,a-l-s/2}; that ELKO fields verify second-order derivative scalar-like field equations is the reason for which they are such an interesting field in cosmology \cite{b/2,b-m,b-b/1,b-b/2,b-b-m-s}, and this is the phenomenological reason for which ELKO spinor fields have attracted much attention. Other properties of ELKOs have been investigated in references \cite{a,a-h} and \cite{r-r,dr-hs,r-h,hs-dr} in terms of their algebraic structure and in \cite{d-dc-hds,dr-b-hds,dr-hds,Bernardini:2012sc} for their dynamical behaviour.

Clearly, the fact that ELKOs are eigenstates of charge-conjugation means that they are neutral in a natural way, whereas Dirac fields are naturally charged and neutrality can be obtained only after the imposition of additional constraints: their different electrodynamic character is certainly the most important fact that discriminates between these two matter fields. However, as a complementary study, it may also be important to know what distinguishes these two fields under the point of view of their response to gravitation, in general, and to gravity with torsion, in particular, and in the past there have been a few attempts to study this type of problem \cite{fabbri/1,fabbri/2,fabbri/3,fabbri,fabbri-vignolo,FABBRI,FV}.

In the present paper, we will review these attempts altogether under a new light.
%%%%%%%%%%%%%%%%%%%%%%%%%%%%%%%%%%%%%%%%%%%%%%%%%%%%%%%%%%%%%%%%%%%%%%%%%%%%%%%%%%%%%%%%%%%%%%%%%%%
%%%%%%%%%%%%%%%%%%%%%%%%%%%%%%%%%%%%%%%%%%%%%%%%%%%%%%%%%%%%%%%%%%%%%%%%%%%%%%%%%%%%%%%%%%%%%%%%%%%
\section{Geometry and Kinematics}

\subsection{Covariance}
All along this presentation, we will consider spacetime to be a $(1+3)$-dimensional manifold possessing a metric and a differential structure \cite{FV}. The metric tensor is given by $g_{\alpha\sigma}$ and the covariant derivative is given by $D_{\mu}$ in terms of the metric-compatible connection $\Gamma^{\mu}_{\alpha\sigma}$ in the most general case: the condition of metric-compatibility $Dg=0$ allows us to decompose these most general covariant derivative and connection in terms of the simplest covariant derivative $\nabla_{\mu}$ and connection $\Lambda^{\alpha}_{\mu\nu}=\frac{1}{2}g^{\rho\alpha}\left(\partial_{\mu}g_{\nu\rho}
+\partial_{\nu}g_{\mu\rho}-\partial_{\rho}g_{\mu\nu}\right)$ according to the form
\begin{eqnarray}
&\Gamma^{\rho}_{\mu\nu}=\Lambda^{\rho}_{\mu\nu}+K^{\rho}_{\mu\nu}
\end{eqnarray}
where $K^{\rho}_{\phantom{\rho}\mu\nu}=\frac{1}{2}\left(Q^{\rho}_{\phantom{\rho}\mu\nu}
+Q_{\mu\nu}^{\phantom{\mu\nu}\rho}+Q_{\nu\mu}^{\phantom{\nu\mu}\rho}\right)$ is called contorsion and it is given by $Q^{\rho}_{\phantom{\rho}\mu\nu}=\Gamma^{\rho}_{\mu\nu}-\Gamma^{\rho}_{\nu\mu}$ called Cartan torsion tensor; so the most general metric-compatible connection consists of two parts, one being a connection entirely written in terms of the metric and the other being a tensor independent on the metric and further decomposable according to
\begin{eqnarray}
&Q_{\rho\mu\nu}=\frac{1}{6}W^{\alpha}\varepsilon_{\alpha\rho\mu\nu}
+\frac{1}{3}\left(V_{\nu}g_{\rho\mu}-V_{\mu}g_{\rho\nu}\right)+T_{\rho\mu\nu}
\end{eqnarray}
where $T_{\rho\mu\nu}=Q_{\rho\mu\nu}-\frac{1}{6}W^{\alpha}\varepsilon_{\alpha\rho\mu\nu}
-\frac{1}{3}\left(V_{\nu}g_{\rho\mu}-V_{\mu}g_{\rho\nu}\right)$ is the non-completely antisymmetric irreducible part of torsion, where $Q_{\rho\mu\nu}\varepsilon^{\rho\mu\nu\alpha}=W^{\alpha}$ is the completely antisymmetric part of torsion and $Q^{\rho}_{\phantom{\rho}\rho\nu}=V_{\nu}$ is the trace part of torsion, all of them being independent on one another. Equivalently, we may employ the vierbein formalism, in which orthonormal tetrad fields $\xi_{a}^{\sigma}$ are introduced so that $g_{\alpha\nu}\xi_{a}^{\alpha}\xi_{b}^{\nu}= \eta_{ab}$ where $\eta_{ab}$ is the Minkowskian matrix while the covariant derivative denoted with $D_{\mu}$ is defined through the spin-connection $\Gamma^{i}_{j\mu}$ in general: the conditions of compatibility are now expressed in terms of the pair of conditions $D\xi=0$ and $D\eta=0$ respectively equivalent to asking that the connection and spin-connection be related according to the relationship given by
\begin{eqnarray}
&\Gamma^{b}_{\phantom{b}j\mu}\!=\!\xi^{\alpha}_{j}\xi_{\rho}^{b} (\Gamma^{\rho}_{\phantom{\rho}\alpha\mu}+\xi_{\alpha}^{k}\partial_{\mu}\xi^{\rho}_{k})
\end{eqnarray}
and that $\Gamma^{bj}_{\phantom{bj}\mu}\!=\!-\Gamma^{jb}_{\phantom{jb}\mu}$ and again, the first relationship showing that although we cannot define a torsion for the spin-connection nevertheless we may write $-Q^{a}_{\phantom{a}\mu\nu}=\partial_{\mu}\xi^{a}_{\nu}-\partial_{\nu}\xi^{a}_{\mu}
+\Gamma^{a}_{j\mu}\xi^{j}_{\nu}-\Gamma^{a}_{j\nu}\xi^{j}_{\mu}$ while the second relationship representing a property of the spin-connection that is important since we aim to describe the underlying Lorentz structure in complex representation. Before proceeding, we recall that in general complex fields are subject to a phase transformation that will require the introduction of the gauge-covariant derivative $D_{\mu}$ defined through the gauge-connection $A_{\mu}$ as it is usually done in gauge theories. Of all complex representations of the Lorentz group we will be interested in the $\frac{1}{2}$-spin representation, the simplest possible one. For it to be obtained, we introduce the $\boldsymbol{\gamma}_{a}$ matrices belonging to the Clifford algebra $\{\boldsymbol{\gamma}_{i}, \boldsymbol{\gamma}_{j}\}=2\boldsymbol{\mathbb{I}}\eta_{ij}$ from which $\boldsymbol{\sigma}_{ij}\!=\!\frac{1}{4}[\boldsymbol{\gamma}_{i},\boldsymbol{\gamma}_{j}]$ are the antisymmetric matrices belonging to the Lorentz algebra, so that we may take them as the antisymmetric complex generators of the infinitesimal Lorentz complex transformation, called spinorial transformation, for which the spinorial covariant derivative $\boldsymbol{D}_{\mu}$ is defined by the spinorial connection $\boldsymbol{\Gamma}_{\mu}$ in general: conditions $\boldsymbol{D}_{\mu}\boldsymbol{\gamma}_{j}=0$ can be equivalently translated into the fact that the most general spinorial connection can be decomposed in two contributions according to
\begin{eqnarray}
&\boldsymbol{\Gamma}_{\mu}=\frac{1}{2}\Gamma^{ab}_{\phantom{ab}\mu}\boldsymbol{\sigma}_{ab} +ieA_{\mu}\boldsymbol{\mathbb{I}}
\end{eqnarray}
in terms of a Lorentz-valued spin-connection plus an abelian term; we may now see that the antisymmetry of the spin-connection is perfectly matched by the antisymmetry of the generators of the infinitesimal spinorial transformation as anticipated and that the abelian field can be taken as the gauge-connection potential. It is intriguing that the most general spinorial connection contains the tetrads as well as one abelian gauge field; in the interpretation that will naturally follow, the tetrads are what will contain the gravitational information whereas the abelian gauge field will represent electrodynamics and the label $e$ the electric charge. As it is clear, there is yet no indication as to what the mass will actually be. Relationship $\{\boldsymbol{\gamma}_{i}, \boldsymbol{\sigma}_{jk}\}= i\varepsilon_{ijkq} \boldsymbol{\pi}\boldsymbol{\gamma}^{q}$ implicitly defines the $\boldsymbol{\pi}$ matrix\footnote{This matrix $\boldsymbol{\pi}$ is what usually in the literature is indicated with $\boldsymbol{\gamma}^{5}$, but we object to such a notation for two reasons: first the notation was introduced in the context of Kaluza-Klein $5$-dimensional models where the index $5$ represented a fifth coordinate that is no longer meaningful in the $4$-dimensional context, and then that notation seems to suggest that with $\boldsymbol{\gamma}^{5}$ there should be an additional $\boldsymbol{\gamma}_{5}$ and in some convention the two differ by a sign that constitutes yet another convention and a source of confusion; hence in this paper we will simply drop the potentially confusing and anyway useless index $5$ and we will take for the notation what we have given above.} which is used to construct the left-handed and right-handed chiral projectors needed to split the spinor field according to the expressions
\begin{eqnarray}
&\frac{1}{2}(\mathbb{I}-\boldsymbol{\pi})\psi=L\ \ \ \ \ \ \ \ 
\frac{1}{2}(\mathbb{I}+\boldsymbol{\pi})\psi=R
\end{eqnarray}
where $L$ and $R$ are the left-handed and right-handed components, which are irreducible and thus independent semi-spinor fields; by noticing that with the left-handed semi-spinor field $L$ one can construct $\boldsymbol{\sigma}^{2}L^{*}$ which transforms as a right-handed semi-spinor field, then it is possible to see that the column of two semi-spinor fields given by
\begin{eqnarray}
\begin{tabular}{c}
$\lambda=\left(\begin{array}{c}L\\ -\boldsymbol{\sigma}^{2}L^{*}
\end{array}\right)$
\end{tabular}
\end{eqnarray}
has the transformation properties of a spinor and it is what will be the ELKO spinor field: it is important to notice that with such a definition the Dirac field has two independent components whereas the ELKO field has two components that are not independent and thus to recover the correct amount of degrees of freedom it is necessary to notice that the definition of the ELKO field actually comes with a two-fold multiplicity that can be expressed as
\begin{eqnarray}
\begin{tabular}{c}
$\lambda_{+-}=\left(\begin{array}{c}L_{+}\\ -\boldsymbol{\sigma}^{2}L_{+}^{*}
\end{array}\right)$
\end{tabular}\ \ \ \ \ \ \ \ 
\begin{tabular}{c}
$\lambda_{-+}=\left(\begin{array}{c}L_{-}\\ -\boldsymbol{\sigma}^{2}L_{-}^{*}
\end{array}\right)$
\end{tabular}
\end{eqnarray}
whenever they are eigenstates with positive or negative eigenvalue of the helicity operator; the Dirac field is
\begin{eqnarray}
\begin{tabular}{c}
$\psi=\left(\begin{array}{c}L\\ R
\end{array}\right)$
\end{tabular}
\end{eqnarray}
condensing the two helicities within a single spinor field. To define the adjoint procedure for the Dirac field we retain the relationship $\overline{\psi}=\psi^{\dagger}\boldsymbol{\gamma}^{0}$ as usually; for ELKO fields the alternative relationship $\stackrel{\neg}{\lambda}_{\mp\pm}=\pm i\lambda_{\pm\mp}^{\dagger}\boldsymbol{\gamma}^{0}$ must be assigned instead. Hence, Dirac spinors and their adjoint are eigenstates of parity while ELKO spinors and their adjoint are eigenstates of the charge-conjugation operator. It is also worth recalling that spinor fields can be categorized into the Lounesto classification, in which Dirac spinors find place according to their properties in the first, second, third and sixth class, while ELKO spinors find place in the fifth class, but there is also another type of spinor called flag-dipole and which constitutes the content of the fourth class \cite{Vignolo:2011qt,daRocha:2013qhu}. In this paper we will not consider this type-$4$ spinor field, although clearly the analysis we will carry on can be extended as to include spinors of this last type.

In terms of the connection alone it is possible to define the Riemann curvature tensor as
\begin{eqnarray}
G^{\rho}_{\phantom{\rho}\eta\mu\nu}
=\partial_{\mu}\Gamma^{\rho}_{\eta\nu}-\partial_{\nu}\Gamma^{\rho}_{\eta\mu}
+\Gamma^{\rho}_{\sigma\mu}\Gamma^{\sigma}_{\eta\nu}
-\Gamma^{\rho}_{\sigma\nu}\Gamma^{\sigma}_{\eta\mu}
\end{eqnarray}
which in the form $G_{\sigma\eta\rho\nu}$ is antisymmetric in both the first and second couple of indices, so with one independent contraction $G^{\rho}_{\phantom{\rho}\eta\rho\nu}\!=\!G_{\eta\nu}$ with contraction $G_{\eta\nu}g^{\eta\nu}\!=\!G$ called Ricci tensor and scalar, and decomposable according to the relationship $G^{\rho}_{\phantom{\rho}\eta\mu\nu}=R^{\rho}_{\phantom{\rho}\eta\mu\nu}
+\nabla_{\mu}K^{\rho}_{\eta\nu}-\nabla_{\nu}K^{\rho}_{\eta\mu}
+K^{\rho}_{\sigma\mu}K^{\sigma}_{\eta\nu}-K^{\rho}_{\sigma\nu}K^{\sigma}_{\eta\mu}$ in terms of the torsionless curvature given by the Riemann metric curvature tensor $R^{\rho}_{\phantom{\rho}\eta\mu\nu}=\partial_{\mu}\Lambda^{\rho}_{\eta\nu}
-\partial_{\nu}\Lambda^{\rho}_{\eta\mu}+\Lambda^{\rho}_{\sigma\mu}\Lambda^{\sigma}_{\eta\nu}
-\Lambda^{\rho}_{\sigma\nu}\Lambda^{\sigma}_{\eta\mu}$ with contraction $R_{\eta\nu}$ contracted as $R$ called Ricci metric curvature tensor and scalar, as usual; in tetrad formalism curvatures are translated into
\begin{eqnarray}
&G^{a}_{\phantom{a}b\sigma\pi}
=\partial_{\sigma}\Gamma^{a}_{b\pi}-\partial_{\pi}\Gamma^{a}_{b\sigma}
+\Gamma^{a}_{j\sigma}\Gamma^{j}_{b\pi}-\Gamma^{a}_{j\pi}\Gamma^{j}_{b\sigma}
\end{eqnarray}
that is $G^{\rho}_{\phantom{\rho}\eta\mu\nu}\xi^{a}_{\rho}\xi^{\eta}_{b} =G^{a}_{\phantom{a}b\mu\nu}$ as it is to be expected. From the gauge-connection we can define
\begin{eqnarray}
&F_{\mu\nu}=\partial_{\mu}A_{\nu}-\partial_{\nu}A_{\mu}
\end{eqnarray}
as the strength of the gauge-connection potential. Finally, we have that we may define
\begin{eqnarray}
&\boldsymbol{G}_{\mu\nu}=\partial_{\mu}\boldsymbol{\Gamma}_{\nu}
-\partial_{\nu}\boldsymbol{\Gamma}_{\mu}+[\boldsymbol{\Gamma}_{\mu},\boldsymbol{\Gamma}_{\nu}]
\end{eqnarray}
in general; as a consequence, we have that $\boldsymbol{G}_{\mu\nu}
=\frac{1}{2}G^{ab}_{\phantom{ab}\mu\nu}\boldsymbol{\sigma}_{ab} +ieF_{\mu\nu}\boldsymbol{\mathbb{I}}$ again in general. It is important to notice that expressions $[\boldsymbol{D}_{\rho},\boldsymbol{D}_{\mu}]\psi= Q^{\theta}_{\phantom{\theta}\rho\mu}\boldsymbol{D}_{\theta}\psi
+\boldsymbol{G}_{\rho\mu}\psi$ and $[\boldsymbol{D}_{\rho},\boldsymbol{D}_{\mu}]\lambda =Q^{\theta}_{\phantom{\theta}\rho\mu}\boldsymbol{D}_{\theta}\lambda
+\boldsymbol{G}_{\rho\mu}\lambda$ specify that both Dirac and ELKO fields have the same covariant derivative commutator; the the commutator of covariant derivatives shows that torsion enters in addition to curvature in making non-trivial the covariant displacements around a given point of the spacetime continuum: specifically, torsion acts as the covariant generator of the translations while curvature is as usual the covariant generator of the rotations, as first discussed by Sciama and Kibble. In this sense, torsion and curvature can respectively be interpreted as the strength of the potentials that arise when gauging translations and rotations in a gauge theory of the full Poincar\'{e} group; as a consequence, it will be expected to see that torsion and curvature will couple to the spin and energy density of matter fields. All such results have been reviewed in \cite{h-h-k-n}.
%%%%%%%%%%%%%%%%%%%%%%%%%%%%%%%%%%%%%%%%%%%%%%%%%%%%%%%%%%%%%%%%%%%%%%%%%%%%%%%%%%%%%%%%%%%%%%%%%%%
\subsection{Conformal Covariance}
In what we intend to do, it is essential that we also introduce the concept of conformal invariance: the idea of a theory that is invariant under conformal transformation was first discussed by Weyl, who noticed that the conformal dynamics, that is the dynamics that is invariant under conformal transformations, is uniquely defined; apart from this issue of formal elegance, the importance of the conformal dynamics is also increased by Stelle demonstration that conformal gravitation is a renormalizable theory of gravity \cite{s}. Conformal transformations are given in terms of a unique function $\sigma$ and setting $\ln{\sigma}=\phi$ by definition: the most general conformal transformation for the metric is
\begin{eqnarray}
&g_{\alpha\beta}\rightarrow\sigma^{2}g_{\alpha\beta}
\end{eqnarray}
and it is also possible to remark that the transformation given by the law
\begin{eqnarray}
&Q^{\sigma}_{\phantom{\sigma}\rho\alpha}\rightarrow Q^{\sigma}_{\phantom{\sigma}\rho\alpha}
+q(\delta^{\sigma}_{\rho}\partial_{\alpha}\phi-\delta^{\sigma}_{\alpha}\partial_{\rho}\phi)
\label{conformaltorsion}
\end{eqnarray}
can be reasonably seen as a conformal transformation for torsion, since its effects on the connection are analogous to those induced by the conformal transformation of the metric itself; in fact for $q=1$ the two effects of the conformal transformation perfectly cancel one another, although here we will retain the conformal weight $q$ to be the most general possible. As usual, the gauge-connection has no conformal transformation. Conformal properties for the spinor fields are assigned according to the mass dimension that is needed in order to have the kinetic term in the Lagrangian relevant at all scales, and so the choice of conformal weight for spinors will not be justified till the spinorial dynamics is assigned: as usual Dirac fields have mass dimension $\frac{3}{2}$ while ELKO fields have mass dimension $1$
\begin{eqnarray}
&\psi\rightarrow\sigma^{-\frac{3}{2}}\psi\ \ \ \ \ \ \ \ 
\mathrm{and}\ \ \ \ \ \ \ \ \lambda\rightarrow\sigma^{-1}\lambda
\end{eqnarray}
as we had implicitly anticipated in the introduction when saying that ELKO fields have to be characterized by a Klein-Gordon type of field equations. As we have already stressed, for the moment the reader will have to take these assignments for granted, although they will be justified when the spinorial field equations will be given later on.

The problem of allowing conformal transformations of torsion, whether weak or strong, is that because of the presence of torsionless derivatives of torsion, there will always be a term with a conformal transformation that spoils the conformal transformation of the irreducible part of the Riemann curvature: so while the torsionless tensor
\begin{eqnarray}
&C_{\alpha\beta\mu\nu}=R_{\alpha\beta\mu\nu}
-\frac{1}{2}(R_{\alpha[\mu}g_{\nu]\beta}-R_{\beta[\mu}g_{\nu]\alpha})
+\frac{1}{12}R(g_{\alpha[\mu}g_{\nu]\beta}-g_{\beta[\mu}g_{\nu]\alpha})
\end{eqnarray}
is conformally covariant the tensor $W_{\alpha\beta\mu\nu}=G_{\alpha\beta\mu\nu}
-\frac{1}{2}(G_{\alpha[\mu}g_{\nu]\beta}-G_{\beta[\mu}g_{\nu]\alpha})
+\frac{1}{12}G(g_{\alpha[\mu}g_{\nu]\beta}-g_{\beta[\mu}g_{\nu]\alpha})$ will never be unless some additional term is used to compensate the spurious terms arising from torsion, thus recovering the conformal invariance in presence of torsion: one such term can be obtained in terms of the modified curvature
\begin{eqnarray}
&M_{\alpha\beta\mu\nu}=G_{\alpha\beta\mu\nu}
+(\frac{1-q}{3q})(V_{\beta}Q_{\alpha\mu\nu}-V_{\alpha}Q_{\beta\mu\nu})
\label{curvature}
\end{eqnarray}
which is such that its irreducible part
\begin{eqnarray}
&T_{\alpha\beta\mu\nu}=M_{\alpha\beta\mu\nu}
-\frac{1}{2}(M_{\alpha[\mu}g_{\nu]\beta}-M_{\beta[\mu}g_{\nu]\alpha})
+\frac{1}{12}M(g_{\alpha[\mu}g_{\nu]\beta}-g_{\beta[\mu}g_{\nu]\alpha})
\label{conformalcurvature}
\end{eqnarray}
is conformally covariant. We have to specify that the conformal transformation for torsion (\ref{conformaltorsion}) is the most general that can be given, including the special case $q=0$ in which torsion has no conformal transformation at all: the two cases represented by the conformal weights $q=0$ and $q\neq0$ are respectively called weak and strong conformal transformations, as in the discussion by Shapiro in \cite{sh}, but such two cases will have to be treated separately, since only strong conformal transformations can be used in (\ref{curvature}) and therefore in (\ref{conformalcurvature}); furthermore, it is also worth remarking that despite (\ref{curvature}) is such that (\ref{conformalcurvature}) is conformally covariant, nevertheless this choice has never been proven to be unique, and thus alternative models might be possible. In \cite{f} and \cite{F} we have studied respectively the case of weak conformal transformations and strong conformal transformations with the special choice (\ref{curvature}-\ref{conformalcurvature}); in this paper we will focusing only on strong conformal transformations with the conformal tensor (\ref{conformalcurvature}). More general cases will be characterized by such a complexity, that for our purposes the model we will take into account here is good enough.

In this geometric background where kinematic quantities are defined we will next set the dynamical action.
%%%%%%%%%%%%%%%%%%%%%%%%%%%%%%%%%%%%%%%%%%%%%%%%%%%%%%%%%%%%%%%%%%%%%%%%%%%%%%%%%%%%%%%%%%%%%%%%%%%
%%%%%%%%%%%%%%%%%%%%%%%%%%%%%%%%%%%%%%%%%%%%%%%%%%%%%%%%%%%%%%%%%%%%%%%%%%%%%%%%%%%%%%%%%%%%%%%%%%%
\section{Dynamics}

\subsection{Covariance}
With the kinematic quantities we have defined, next task will be that of setting up the dynamical Lagrangian that will determine the field equations of the fields, and a principle we will consider in order to do that will be the requirement of taking everything at the least-order derivative possible; there is no real reason as to why this has to be done but one of simplicity, and that is if this were not the case, then there will be no principle that could prevent us to take whatever term of whichever derivative order, the action will be featured by an infinite number of constants and the theory would lose any predictive power: to avoid this, we will take least-order derivative terms.

To look for the least-order derivative Lagrangian in general, we have to notice that a Lagrangian that is least-order derivative can only be linear in the curvature and quadratic in torsion: as an easy computation shows, for a linear expression of curvature we can construct only the Ricci scalar $G$, and for a quadratic expression of torsion we can construct the three independent scalars $Q^{\nu}_{\phantom{\nu}\nu\mu}Q_{\rho}^{\phantom{\rho}\rho\mu}$, $Q_{\rho\mu\nu}Q^{\rho\mu\nu}$, $Q_{\rho\mu\nu}Q^{\nu\mu\rho}$ or equivalently, if we decompose the curvature in terms of the metric curvature plus torsion and subsequently if we decompose torsion into its three irreducible parts then we have the Ricci metric curvature scalar $R$, with $V_{\mu}V^{\mu}$, $W_{\rho}W^{\rho}$, $T_{\rho\mu\nu}T^{\rho\mu\nu}$ in general, so that we have that the torsion-gravitational least-order derivative Lagrangian in its most general form is given by
\begin{eqnarray}
&{\cal L}_G=G+AQ^{\nu}_{\phantom{\nu}\nu\mu}Q_{\rho}^{\phantom{\rho}\rho\mu}
+BQ_{\rho\mu\nu}Q^{\rho\mu\nu}+CQ_{\rho\mu\nu}Q^{\nu\mu\rho}
\label{lagrangian}
\end{eqnarray}
in terms of four coupling constants, that is the Newton gravitational coupling constant and the three torsional coupling constants given by $A$, $B$, $C$ independent on one another and yet to be determined; the electrodynamic Lagrangian
\begin{eqnarray}
&{\cal L}_E=-\frac{1}{4}F_{\alpha\beta}F^{\alpha\beta}
\end{eqnarray}
is as usually done. For the matter fields, the Lagrangians are different, as expected: the Dirac field has the usual
\begin{eqnarray}
&{\cal L}_{\textrm{Dirac}}=\frac{i}{2}(\overline{\psi}\boldsymbol{\gamma}^{\mu}\boldsymbol{D}_{\mu}\psi
-\boldsymbol{D}_{\mu}\overline{\psi}\boldsymbol{\gamma}^{\mu}\psi)-m\overline{\psi}\psi
\end{eqnarray}
in terms of the mass $m$ of the matter field; as anticipated, the ELKO Lagrangian is given by the scalar-like Lagrangian
\begin{eqnarray}
{\cal L}_{\textrm{ELKO}}=\boldsymbol{D}_{\alpha}\stackrel{\neg}{\lambda}(\eta^{\alpha\beta}\mathbb{I} +a\boldsymbol{\sigma}^{\alpha\beta})\boldsymbol{D}_{\beta}\lambda
-m^{2}\stackrel{\neg}{\lambda}\lambda
\end{eqnarray}
in terms of the mass $m$ and a free parameter $a$ to be determined. As it is known the Dirac Lagrangian cannot contain more than one derivative and thus the index must be contracted with a $\boldsymbol{\gamma}^{\mu}$ matrix; on the other hand, the ELKO Lagrangian has two derivatives and thus the two indices can be contracted with two $\boldsymbol{\gamma}^{\mu}$ matrices as well as between one another, as for scalar fields, and this gives rise to two possibilities, justifying the fact that the ELKO action has one more free parameter. It is however essential to notice that such additional term does not enter into the action only because ELKO fields have the algebraic features of spinors and the dynamical features of scalars but also because of the presence of torsion, since with no torsion such term would reduce to a divergence and it would be irrelevant.

Upon variation with respect to all independent fields involved, the field equations will be
\begin{eqnarray}
\nonumber
&C\left(Q^{\mu\nu\rho}-Q^{\nu\mu\rho}+2Q^{\rho\mu\nu}\right)
+B\left(2Q^{\nu\mu\rho}-2Q^{\mu\nu\rho}\right)+\\
&+A\left(V^{\nu}g^{\rho\mu}-V^{\mu}g^{\rho\nu}\right)
+\left(Q^{\rho\mu\nu}+V^{\mu}g^{\rho\nu}-V^{\nu}g^{\rho\mu}\right)=-S^{\rho\mu\nu}
\label{Sciama-Kibble}
\end{eqnarray}
for the torsion-spin coupling and
\begin{eqnarray}
\nonumber
&C\left(D_{\mu}Q^{\mu\rho\alpha}-D_{\mu}Q^{\rho\mu\alpha}
+V_{\mu}Q^{\mu\rho\alpha}-V_{\mu}Q^{\rho\mu\alpha} +Q^{\theta\sigma\alpha}Q_{\sigma\theta}^{\phantom{\sigma\theta}\rho}
-\frac{1}{2}Q^{\theta\sigma\pi}Q_{\pi\sigma\theta}g^{\rho\alpha}\right)+\\
\nonumber
&+B\left(2D_{\mu}Q^{\alpha\rho\mu}+2V_{\mu}Q^{\alpha\rho\mu}
+2Q^{\theta\sigma\alpha}Q_{\theta\sigma}^{\phantom{\theta\sigma}\rho}
-Q^{\rho\theta\sigma}Q^{\alpha}_{\phantom{\alpha}\theta\sigma}
-\frac{1}{2}Q^{\theta\sigma\pi}Q_{\theta\sigma\pi}g^{\rho\alpha}\right)+\\
&+A\left(-D^{\alpha}V^{\rho}+D_{\mu}V^{\mu}g^{\rho\alpha}
+\frac{1}{2}V_{\mu}V^{\mu}g^{\rho\alpha}\right)
+\left(G^{\rho\alpha}-\frac{1}{2}Gg^{\rho\alpha}\right)
-\frac{1}{2}\left(\frac{1}{4}g^{\rho\alpha}F^{2}-F^{\rho\theta}F^{\alpha}_{\phantom{\alpha}\theta}\right)
=\frac{1}{2}T^{\rho\alpha}
\label{Einstein}
\end{eqnarray}
for the curvature-energy coupling; there will also be the field equations
\begin{eqnarray}
&\frac{1}{2}F_{\mu\nu}Q^{\rho\mu\nu}+F^{\mu\rho}V_{\mu}+D_{\sigma}F^{\sigma\rho}=eJ^{\rho}
\label{Maxwell}
\end{eqnarray}
for the gauge-current coupling. It is possible to prove that some geometrical identities known as Bianchi identities can be used to demonstrate that these field equations are converted into the following expressions
\begin{eqnarray}
&D_{\rho}S^{\rho\mu\nu}+V_{\rho}S^{\rho\mu\nu}
+\frac{1}{2}T^{[\mu\nu]}=0
\label{spincons}\\
&D_{\mu}T^{\mu\rho}+V_{\mu}T^{\mu\rho}-T_{\mu\sigma}Q^{\sigma\mu\rho}
+S_{\beta\mu\sigma}G^{\sigma\mu\beta\rho}+eJ_{\beta}F^{\beta\rho}=0
\label{energycons}
\end{eqnarray}
and also
\begin{eqnarray}
D_{\rho}J^{\rho}+V_{\rho}J^{\rho}=0
\label{gaugecons}
\end{eqnarray}
as conservation laws for the spin $S^{\rho\mu\nu}$ and energy $T^{\mu\rho}$ and for the current $J^{\rho}$ that have to be verified by these quantities, once the matter field equations are assigned. Such conserved quantities are given alongside to the matter field equations by the matter Lagrangian, and given that we are comparing two different matter fields, correspondingly we will have two different sets of conserved quantities and matter field equations: for the Dirac field we have the completely antisymmetric spin and the non-symmetric energy given by the following expressions
\begin{eqnarray}
&S_{\mu\alpha\beta}=
\frac{i}{4}\overline{\psi}\{\boldsymbol{\gamma}_{\mu},\boldsymbol{\sigma}_{\alpha\beta}\}\psi
\label{Diracspin}\\
&T_{\mu\alpha}=\frac{i}{2}(\overline{\psi}\boldsymbol{\gamma}_{\mu}\boldsymbol{D}_{\alpha}\psi-\boldsymbol{D}_{\alpha}\overline{\psi}
\boldsymbol{\gamma}_{\mu}\psi)
\label{Diracenergy}
\end{eqnarray}
together with the current
\begin{eqnarray}
&J_{\mu}=\overline{\psi}\boldsymbol{\gamma}_{\mu}\psi
\label{Diraccurrent}
\end{eqnarray}
as usual, and we have of course the Dirac field equations
\begin{eqnarray}
&i\boldsymbol{\gamma}^{\mu}\boldsymbol{D}_{\mu}\psi
+\frac{i}{2}\boldsymbol{\gamma}^{\mu}V_{\mu}\psi-m\psi=0
\label{Diracmatter}
\end{eqnarray}
as they are widely known; for the ELKO fields we have the spin and the symmetric energy
\begin{eqnarray}
&S_{\mu\alpha\beta}
=\frac{1}{2}\left(\boldsymbol{D}_{\mu}\stackrel{\neg}{\lambda}
\boldsymbol{\sigma}_{\alpha\beta}\lambda
-\stackrel{\neg}{\lambda}\boldsymbol{\sigma}_{\alpha\beta}
\boldsymbol{D}_{\mu}\lambda\right)
+\frac{a}{2}\left(\boldsymbol{D}^{\rho}\stackrel{\neg}{\lambda}
\boldsymbol{\sigma}_{\rho\mu}\boldsymbol{\sigma}_{\alpha\beta}\lambda
-\stackrel{\neg}{\lambda}\boldsymbol{\sigma}_{\alpha\beta}
\boldsymbol{\sigma}_{\mu\rho}\boldsymbol{D}^{\rho}\lambda\right)
\label{ELKOspin}\\
\nonumber
&T_{\mu\nu}=
\left(\boldsymbol{D}_{\mu}\stackrel{\neg}{\lambda}\boldsymbol{D}_{\nu}\lambda
+\boldsymbol{D}_{\nu}\stackrel{\neg}{\lambda}\boldsymbol{D}_{\mu}\lambda
-g_{\mu\nu}\boldsymbol{D}_{\rho}\stackrel{\neg}{\lambda}\boldsymbol{D}^{\rho}\lambda\right)+\\
&+a\left(\boldsymbol{D}_{\nu}\stackrel{\neg}{\lambda}
\boldsymbol{\sigma}_{\mu\rho}\boldsymbol{D}^{\rho}\lambda
+\boldsymbol{D}^{\rho}\stackrel{\neg}{\lambda}
\boldsymbol{\sigma}_{\rho\mu}\boldsymbol{D}_{\nu}\lambda
-g_{\mu\nu}\boldsymbol{D}_{\rho}\stackrel{\neg}{\lambda}
\boldsymbol{\sigma}^{\rho\sigma}\boldsymbol{D}_{\sigma}\lambda\right)
+g_{\mu\nu}m^{2}\stackrel{\neg}{\lambda}\lambda
\label{ELKOenergy}
\end{eqnarray}
with no current since ELKO fields are neutral, and we have the ELKO field equations
\begin{eqnarray}
&\left(\boldsymbol{D}^{2}\lambda+V^{\mu}\boldsymbol{D}_{\mu}\lambda\right)
+a\left(\boldsymbol{\sigma}^{\rho\mu}\boldsymbol{D}_{\rho}\boldsymbol{D}_{\mu}\lambda
+V_{\rho}\boldsymbol{\sigma}^{\rho\mu}\boldsymbol{D}_{\mu}\lambda\right)
+m^{2}\lambda=0
\label{ELKOmatter}
\end{eqnarray}
as the scalar-type of field equations we anticipated. We notice that the conservation laws (\ref{spincons}-\ref{energycons}) and (\ref{gaugecons}) are in fact verified when the matter field equations are satisfied, and of course this is true for both the Dirac and the ELKO models; when all is put together, the whole set of field equations for the Dirac system is
\begin{eqnarray}
&\frac{1}{3}(4C-4B+1)W^{\sigma}=\frac{1}{2}
\overline{\psi}\boldsymbol{\gamma}^{\sigma}\boldsymbol{\pi}\psi
\label{torsion-spinDirac}\\
\nonumber
&\frac{1}{3}\left(B-C\right)\left(D_{\mu}W_{\sigma}\varepsilon^{\mu\sigma\rho\alpha}
+\frac{1}{6}W^{\alpha}W^{\rho}+\frac{1}{12}W^{2}g^{\alpha\rho}\right)
+\left(G^{\rho\alpha}-\frac{1}{2}Gg^{\rho\alpha}\right)-\\
&-\frac{1}{2}\left(\frac{1}{4}g^{\rho\alpha}F^{2}-F^{\rho\theta}F^{\alpha}_{\phantom{\alpha}\theta}\right)
=\frac{i}{4}(\overline{\psi}\boldsymbol{\gamma}^{\rho}\boldsymbol{D}^{\alpha}\psi
-\boldsymbol{D}^{\alpha}\overline{\psi}\boldsymbol{\gamma}^{\rho}\psi)
\label{curvature-energyDirac}
\end{eqnarray}
and
\begin{eqnarray}
&\frac{1}{12}F_{\mu\nu}W_{\sigma}\varepsilon^{\mu\nu\sigma\rho}+D_{\sigma}F^{\sigma\rho}
=e\overline{\psi}\boldsymbol{\gamma}^{\rho}\psi
\label{gauge-currentDirac}
\end{eqnarray}
with
\begin{equation}
i\boldsymbol{\gamma}^{\mu}\boldsymbol{D}_{\mu}\psi-m\psi=0
\label{matterDirac}
\end{equation}
while the whole set of field equations for the ELKO system is
\begin{eqnarray}
\nonumber
&-2C\left(Q^{\mu\nu\rho}-Q^{\nu\mu\rho}+2Q^{\rho\mu\nu}\right)
-2B\left(2Q^{\nu\mu\rho}-2Q^{\mu\nu\rho}\right)
-2A\left(V^{\nu}g^{\rho\mu}-V^{\mu}g^{\rho\nu}\right)+\\
&-2\left(Q^{\rho\mu\nu}+V^{\mu}g^{\rho\nu}-V^{\nu}g^{\rho\mu}\right)
=\left(\boldsymbol{D}^{\rho}\stackrel{\neg}{\lambda}\boldsymbol{\sigma}^{\mu\nu}\lambda
-\stackrel{\neg}{\lambda}\boldsymbol{\sigma}^{\mu\nu}\boldsymbol{D}^{\rho}\lambda\right)
+a\left(\boldsymbol{D}_{\pi}\stackrel{\neg}{\lambda}\boldsymbol{\sigma}^{\pi\rho}
\boldsymbol{\sigma}^{\mu\nu}\lambda-\stackrel{\neg}{\lambda}\boldsymbol{\sigma}^{\mu\nu}
\boldsymbol{\sigma}^{\rho\pi}\boldsymbol{D}_{\pi}\lambda\right)
\label{torsion-spinELKO}\\
\nonumber
&2C\left(D_{\mu}Q^{\mu\rho\alpha}-D_{\mu}Q^{\rho\mu\alpha}
+V_{\mu}Q^{\mu\rho\alpha}-V_{\mu}Q^{\rho\mu\alpha} +Q^{\theta\sigma\alpha}Q_{\sigma\theta}^{\phantom{\sigma\theta}\rho}
-\frac{1}{2}Q^{\theta\sigma\pi}Q_{\pi\sigma\theta}g^{\rho\alpha}\right)+\\
\nonumber
&+2B\left(2D_{\mu}Q^{\alpha\rho\mu}+2V_{\mu}Q^{\alpha\rho\mu}
+2Q^{\theta\sigma\alpha}Q_{\theta\sigma}^{\phantom{\theta\sigma}\rho}
-Q^{\rho\theta\sigma}Q^{\alpha}_{\phantom{\alpha}\theta\sigma}
-\frac{1}{2}Q^{\theta\sigma\pi}Q_{\theta\sigma\pi}g^{\rho\alpha}\right)+\\
\nonumber
&+2A\left(-D^{\alpha}V^{\rho}+D_{\mu}V^{\mu}g^{\rho\alpha}
+\frac{1}{2}V_{\mu}V^{\mu}g^{\rho\alpha}\right)+\\
\nonumber
&+\left(2G^{\rho\alpha}-Gg^{\rho\alpha}\right)
=\left(\boldsymbol{D}^{\rho}\stackrel{\neg}{\lambda}\boldsymbol{D}^{\alpha}\lambda
+\boldsymbol{D}^{\alpha}\stackrel{\neg}{\lambda}\boldsymbol{D}^{\rho}\lambda
-g^{\rho\alpha}\boldsymbol{D}_{\pi}\stackrel{\neg}{\lambda}\boldsymbol{D}^{\pi}\lambda\right)+\\
&+a\left(\boldsymbol{D}^{\alpha}\stackrel{\neg}{\lambda}
\boldsymbol{\sigma}^{\rho\pi}\boldsymbol{D}_{\pi}\lambda
+\boldsymbol{D}_{\pi}\stackrel{\neg}{\lambda}
\boldsymbol{\sigma}^{\pi\rho}\boldsymbol{D}^{\alpha}\lambda
-g^{\rho\alpha}\boldsymbol{D}_{\pi}\stackrel{\neg}{\lambda}
\boldsymbol{\sigma}^{\pi\sigma}\boldsymbol{D}_{\sigma}\lambda\right)
+g^{\rho\alpha}m^{2}\stackrel{\neg}{\lambda}\lambda
\label{curvature-energyELKO}
\end{eqnarray}
with
\begin{eqnarray}
&\left(\boldsymbol{D}^{2}\lambda+V^{\mu}\boldsymbol{D}_{\mu}\lambda\right)
+a\left(\boldsymbol{\sigma}^{\rho\mu}\boldsymbol{D}_{\rho}\boldsymbol{D}_{\mu}\lambda
+V_{\rho}\boldsymbol{\sigma}^{\rho\mu}\boldsymbol{D}_{\mu}\lambda\right)+m^{2}\lambda=0
\label{matterELKO}
\end{eqnarray}
as it is straightforward to check by substitution. It is to be noticed that in the case of the Dirac field because of the algebraic form of the torsion-spin field equations the complete antisymmetry of the spin imposes the complete antisymmetry of torsion while the same cannot be said for ELKO fields since their spin has no particular symmetry and thus torsion has no particular symmetry too, and this is the reason why for the Dirac field we have been employing the completely antisymmetric part of the torsion tensor while for ELKO field torsion has been left general.

Another most important thing we have to notice about the spin-torsion field equations is that in the case in which the spin density vanishes then torsion vanish, and therefore the Newtonian limit is recovered; when in general the torsion and the spin do not vanish the torsion-spin field equations provide the mean with which to have torsion substituted in terms of the spin of the spinorial fields. Again, we have two cases: for the Dirac system, such a substitution is straightforward leading to the symmetrized gravitational field equations written for the Ricci tensor
\begin{eqnarray}
&R^{\rho\alpha}-\frac{1}{2}\left(\frac{1}{4}g^{\rho\alpha}F^{2}-F^{\rho\theta}F^{\alpha}_{\phantom{\alpha}\theta}\right)
=\frac{i}{8}\left(\overline{\psi}\boldsymbol{\gamma}^{\rho}\boldsymbol{\nabla}^{\alpha}\psi
-\boldsymbol{\nabla}^{\alpha}\overline{\psi}\boldsymbol{\gamma}^{\rho}\psi
+\overline{\psi}\boldsymbol{\gamma}^{\alpha}\boldsymbol{\nabla}^{\rho}\psi
-\boldsymbol{\nabla}^{\rho}\overline{\psi}\boldsymbol{\gamma}^{\alpha}\psi\right)
-\frac{1}{4}g^{\rho\alpha}m\overline{\psi}\psi
\end{eqnarray}
and electrodynamic field equations
\begin{eqnarray}
&\nabla_{\sigma}F^{\sigma\rho}=e\overline{\psi}\boldsymbol{\gamma}^{\rho}\psi
\end{eqnarray}
with the matter field equations
\begin{eqnarray}
&i\boldsymbol{\gamma}^{\mu}\boldsymbol{\nabla}_{\mu}\psi
+\frac{3}{16(4C-4B+1)}\overline{\psi}\boldsymbol{\gamma}^{\mu}\boldsymbol{\pi}\psi
\boldsymbol{\gamma}_{\mu}\boldsymbol{\pi}\psi-m\psi=0
\end{eqnarray}
showing that the gravitational and electrodynamic field equations are formally those we would have in the torsionless case while the matter field equations become formally identical to those we would have had in the torsionless case with additional potentials of self-interactions having a yet undetermined coupling constant and with the widely known Nambu--Jona-Lasinio structure; for the ELKO system however such a substitution is not at all easy because we do not have algebraic torsion-spin field equations in the first place. The reason of this fact is simple: like the Dirac field the ELKO field is a spinor and so it has a spin tensor, but differently from Dirac fields the ELKO fields satisfy second-order field equations and as such their spin has a derivative, which itself contains torsion. The ELKO spin-torsion field equations are equations in which torsion is related to the derivatives of the spinor themselves with torsion, so that such equations define torsion implicitly, and obtaining an explicit torsion has never been done in general \cite{FV}.
%%%%%%%%%%%%%%%%%%%%%%%%%%%%%%%%%%%%%%%%%%%%%%%%%%%%%%%%%%%%%%%%%%%%%%%%%%%%%%%%%%%%%%%%%%%%%%%%%%%
\subsection{Conformal Covariance}
As we have seen, the principle of requiring least-order derivative field equations was enough to have the action of the system fixed once and for all; on the other hand, this principle will prove to be useless for conformal models since the requirement of conformal symmetry alone is enough to fix the action of the system.

To look for conformal actions, we have to notice that in $4$-dimensional spaces conformal invariance can only be achieved by terms containing two curvatures, that is in our case terms containing two of the conformal tensors given in definition (\ref{conformalcurvature}), and because of the irreducibility of such tensor and its symmetry properties for indices transposition, it is possible to see that there are exactly three possible ways to contract indices in order to obtain conformal scalars and thus the most general conformal scalar is given by $AT^{\alpha\beta\mu\nu}T_{\alpha\beta\mu\nu}\!+\!BT^{\alpha\beta\mu\nu}T_{\mu\nu\alpha\beta}
\!+\!CT^{\alpha\beta\mu\nu}T_{\alpha\mu\beta\nu}$ in terms of the $A$, $B$, $C$ parameters: through these parameters $A$, $B$, $C$, it is possible to defined the parametric quantity
\begin{eqnarray}
&P_{\alpha\beta\mu\nu}=AT_{\alpha\beta\mu\nu}+BT_{\mu\nu\alpha\beta}+\frac{C}{4}(T_{\alpha\mu\beta\nu}-T_{\beta\mu\alpha\nu}+T_{\beta\nu\alpha\mu}-T_{\alpha\nu\beta\mu})
\label{parametricconformalcurvature}
\end{eqnarray}
which is antisymmetric in the first and second pair of indices, irreducible and conformally covariant, so that the above most general invariant reduces to the form given by $T^{\alpha\beta\mu\nu}P_{\alpha\beta\mu\nu}$ and so the most general action is given by
\begin{eqnarray}
&S_G=\int[kT^{\alpha\beta\mu\nu}P_{\alpha\beta\mu\nu}+L_{\mathrm{matter}}]\sqrt{|g|}dV
\end{eqnarray}
with $k$ gravitational constant, so far undetermined. Once again we have two cases: for the Dirac field masslessness is enough to ensure the conformal properties, and thus we have that the action
\begin{eqnarray}
&S_{\textrm{Dirac}}=\int[\frac{i}{2}(\overline{\psi}\boldsymbol{\gamma}^{\rho}\boldsymbol{D}_{\rho}\psi
-\boldsymbol{D}_{\rho}\overline{\psi}\boldsymbol{\gamma}^{\rho}\psi)]|e|dV
\end{eqnarray}
is the Dirac conformal action, which is the one we had before and thus showing that the Dirac field is naturally adapted to conformal symmetry; but the ELKO field has mass dimension equal to that of a scalar field and as a consequence, for ELKO field much like for scalar fields, additional terms must be added to have the conformal symmetry of the whole dynamical action, and when this is done one of the simplest ELKO conformal actions is given by
\begin{eqnarray}
\nonumber
&S_{\textrm{ELKO}}=\int[\boldsymbol{D}_{\rho}\stackrel{\neg}{\lambda}\boldsymbol{D}^{\rho}\lambda
+a\boldsymbol{D}_{\rho}\stackrel{\neg}{\lambda} \boldsymbol{\sigma}^{\rho\beta}\boldsymbol{D}_{\beta}\lambda+\\
\nonumber
&+(\frac{aq+q-1}{3q})
(\boldsymbol{D}_{\alpha}\stackrel{\neg}{\lambda}\boldsymbol{\sigma}^{\alpha\nu}\lambda
-\stackrel{\neg}{\lambda}\boldsymbol{\sigma}^{\alpha\nu}\boldsymbol{D}_{\alpha}\lambda)V_{\nu}
+(\frac{4-3a+3qa-24p+24pq}{12q})\boldsymbol{D}_{\nu}\lambda^{2}V^{\nu}+\\
&+(\frac{7-3a-6q+3aq^{2}+3q^{2}-24p-24pq+48pq^{2}}{36q^{2}})\lambda^{2}V_{\alpha}V^{\alpha}
+p\lambda^{2}M]|e|dV
\label{action}
\end{eqnarray}
in terms of two free parameters $a$ and $p$ to be determined. As it is rather clear, the ELKO dynamical term is not conformally invariant on its own but the same can be said for some interacting terms between ELKO and the background, so that when all are taken into account and a wise fine-tuning is considered, it is possible to have conformal symmetry after all; the Dirac field instead is already conformal. We will discuss this difference in a moment.

By varying the action, we get the conformal torsion-gravitational field equations in the form
\begin{eqnarray}
&4k[D_{\rho}P^{\alpha\beta\mu\rho}+V_{\rho}P^{\alpha\beta\mu\rho}
-\frac{1}{2}Q^{\mu}_{\phantom{\mu}\rho\theta}P^{\alpha\beta\rho\theta}
-(\frac{1-q}{3q})(V_{\rho}P^{\rho[\alpha\beta]\mu}
-\frac{1}{2}Q_{\sigma\rho\theta}g^{\mu[\alpha}P^{\beta]\sigma\rho\theta})]=S^{\mu\alpha\beta}
\label{Fabbri}\\
\nonumber
&2k[P^{\theta\sigma\rho\alpha}T_{\theta\sigma\rho}^{\phantom{\theta\sigma\rho}\mu}
-\frac{1}{4}g^{\alpha\mu}P^{\theta\sigma\rho\beta}T_{\theta\sigma\rho\beta}
+P^{\mu\sigma\alpha\rho}M_{\sigma\rho}
+(\frac{1-q}{3q})(D_{\nu}(2P^{\mu\rho\alpha\nu}V_{\rho}
-g^{\mu\alpha}P^{\nu\theta\rho\sigma}Q_{\theta\rho\sigma}
+g^{\mu\nu}P^{\alpha\theta\rho\sigma}Q_{\theta\rho\sigma})+\\
&+V_{\nu}(2P^{\mu\rho\alpha\nu}V_{\rho}
-g^{\mu\alpha}P^{\nu\theta\rho\sigma}Q_{\theta\rho\sigma}
-P^{\mu\nu\rho\sigma}Q^{\alpha}_{\phantom{\alpha}\rho\sigma}))]=\frac{1}{2}T^{\alpha\mu}
\label{Weyl}
\end{eqnarray}
for the conformal spin-torsion and energy-curvature coupling. The Bianchi identities still convert these field equations into conservation laws but now because of an additional symmetry there will be three conservation laws
\begin{eqnarray}
&D_{\rho}S^{\rho\mu\nu}+V_{\rho}S^{\rho\mu\nu}
+\frac{1}{2}T^{[\mu\nu]}=0
\label{conservationlawspin}\\
&D_{\mu}T^{\mu\rho}+V_{\mu}T^{\mu\rho}-T_{\mu\sigma}Q^{\sigma\mu\rho}
+S_{\beta\mu\sigma}G^{\sigma\mu\beta\rho}=0
\label{conservationlawenergy}\\
&(1-q)(D_{\mu}S_{\nu}^{\phantom{\nu}\nu\mu}+V_{\mu}S_{\nu}^{\phantom{\nu}\nu\mu})
+\frac{1}{2}T_{\mu}^{\phantom{\mu}\mu}=0
\label{conservationlawtrace}
\end{eqnarray}
which will have to be verified, once the conformal matter field equations are satisfied. Let us derive the contributions of the two conformal matter fields: for the Dirac conformal matter field we have that the completely antisymmetric spin and the non-symmetric traceless energy are given by the expected
\begin{eqnarray}
&S_{\mu\alpha\beta}=
\frac{1}{4}\varepsilon_{\mu\alpha\beta\rho}
\overline{\psi}\boldsymbol{\gamma}^{\rho}\boldsymbol{\pi}\psi\\
&T_{\mu\alpha}=\frac{i}{2}(\overline{\psi}\boldsymbol{\gamma}_{\mu}\boldsymbol{D}_{\alpha}\psi
-\boldsymbol{D}_{\alpha}\overline{\psi}\boldsymbol{\gamma}_{\mu}\psi)
\end{eqnarray}
along with the massless matter field equations
\begin{eqnarray}
&i\boldsymbol{\gamma}^{\mu}\boldsymbol{D}_{\mu}\psi
+\frac{i}{2}V_{\mu}\boldsymbol{\gamma}^{\mu}\psi=0
\end{eqnarray}
as a simple calculation would show; but for the ELKO conformal matter field the spin and energy are
\begin{eqnarray}
\nonumber
&S^{\mu\alpha\beta}=
\frac{1}{2}(\boldsymbol{D}^{\mu}\stackrel{\neg}{\lambda}\boldsymbol{\sigma}^{\alpha\beta}\lambda
-\stackrel{\neg}{\lambda}\boldsymbol{\sigma}^{\alpha\beta}\boldsymbol{D}^{\mu}\lambda)
+\frac{a}{2}(\boldsymbol{D}_{\rho}\stackrel{\neg}{\lambda}
\boldsymbol{\sigma}^{\rho\mu}\boldsymbol{\sigma}^{\alpha\beta}\lambda
-\stackrel{\neg}{\lambda}\boldsymbol{\sigma}^{\alpha\beta}\boldsymbol{\sigma}^{\mu\rho}
\boldsymbol{D}_{\rho}\lambda)+\\
\nonumber
&+(\frac{1-q-aq}{6q})(\boldsymbol{D}^{\rho}\stackrel{\neg}{\lambda}
\boldsymbol{\sigma}_{\rho\theta}\lambda
-\stackrel{\neg}{\lambda}\boldsymbol{\sigma}_{\rho\theta}
\boldsymbol{D}^{\rho}\lambda)g^{\theta[\beta}g^{\alpha]\mu}-\\
\nonumber
&-(\frac{4-3a+3aq-24p}{24q})\boldsymbol{D}_{\theta}\lambda^{2}g^{\theta[\beta}g^{\alpha]\mu}
-(\frac{1-q-aq}{6q})V_{\theta}
\stackrel{\neg}{\lambda}\{\boldsymbol{\sigma}^{\theta\mu},\boldsymbol{\sigma}^{\alpha\beta}\}\lambda+\\
&+(\frac{7-3a-6q+3q^{2}+3aq^{2}-22p-29pq+48pq^{2}}{36q^{2}})V^{[\alpha}g^{\beta]\mu}\lambda^{2}
-pQ^{\mu\alpha\beta}\lambda^{2}\\
\nonumber
&T^{\alpha\mu}=
(\boldsymbol{D}^{\mu}\stackrel{\neg}{\lambda}\boldsymbol{D}^{\alpha}\lambda
+\boldsymbol{D}^{\alpha}\stackrel{\neg}{\lambda}\boldsymbol{D}^{\mu}\lambda
-g^{\mu\alpha}\boldsymbol{D}_{\rho}\stackrel{\neg}{\lambda}\boldsymbol{D}^{\rho}\lambda)+\\
\nonumber
&+a(\boldsymbol{D}^{\mu}\stackrel{\neg}{\lambda}
\boldsymbol{\sigma}^{\alpha\rho}\boldsymbol{D}_{\rho}\lambda
+\boldsymbol{D}_{\rho}\stackrel{\neg}{\lambda}
\boldsymbol{\sigma}^{\rho\alpha}\boldsymbol{D}^{\mu}\lambda
-g^{\mu\alpha}\boldsymbol{D}_{\rho}\stackrel{\neg}{\lambda}
\boldsymbol{\sigma}^{\rho\sigma}\boldsymbol{D}_{\sigma}\lambda)+\\
\nonumber
&+(\frac{1-q-aq}{3q})\boldsymbol{D}_{\beta}(\boldsymbol{D}^{\rho}\stackrel{\neg}{\lambda} \boldsymbol{\sigma}_{\rho\nu}\lambda
-\stackrel{\neg}{\lambda}\boldsymbol{\sigma}_{\rho\nu}\boldsymbol{D}^{\rho}\lambda)
(g^{\alpha\nu}g^{\mu\beta}-g^{\alpha\mu}g^{\nu\beta})-\\
\nonumber
&-(\frac{4-3a+3qa-24p+24pq}{12q})\boldsymbol{D}_{\beta}\boldsymbol{D}_{\nu}\lambda^{2}
(g^{\alpha\nu}g^{\mu\beta}-g^{\alpha\mu}g^{\nu\beta})
+(\frac{1-q-aq}{3q})V_{\rho}(\boldsymbol{D}^{\mu}\stackrel{\neg}{\lambda}
\boldsymbol{\sigma}^{\rho\alpha}\lambda
-\stackrel{\neg}{\lambda}\boldsymbol{\sigma}^{\rho\alpha}\boldsymbol{D}^{\mu}\lambda)-\\
\nonumber
&-(\frac{7-3a-6q+3q^{2}+3aq^{2}-24p+24pq^{2}}{18q^{2}})\boldsymbol{D}_{\beta}(\lambda^{2}V_{\nu})
(g^{\beta\mu}g^{\alpha\nu}-g^{\alpha\mu}g^{\beta\nu})
+(\frac{4-3a+3qa-24p+24pq}{12q})V^{\alpha}\boldsymbol{D}^{\mu}\lambda^{2}+\\
&+(\frac{7-3a-6q+3q^{2}+3aq^{2}-24p+24pq}{36q^{2}})V_{\rho}V^{\rho}\lambda^{2}g^{\alpha\mu}
-\frac{2p(1-q)}{3q}(V^{\alpha}V^{\mu}+Q^{\alpha\mu\rho}V_{\rho})\lambda^{2}
+2p(M^{\alpha\mu}-\frac{1}{2}g^{\alpha\mu}M)\lambda^{2}
\end{eqnarray}
along with the matter field equations
\begin{eqnarray}
\nonumber
&\boldsymbol{D}^{2}\lambda+V^{\mu}\boldsymbol{D}_{\mu}\lambda
+a\boldsymbol{\sigma}^{\rho\mu}\boldsymbol{D}_{\rho}\boldsymbol{D}_{\mu}\lambda
+(\frac{aq-2q+2}{3q})V_{\rho}\boldsymbol{\sigma}^{\rho\mu}\boldsymbol{D}_{\mu}\lambda
+(\frac{aq+q-1}{3q})D_{\rho}V_{\mu}\boldsymbol{\sigma}^{\rho\mu}\lambda
+(\frac{4-3a+3qa-24p+24pq}{12q})D_{\rho}V^{\rho}\lambda-\\
&-(\frac{7-3a-18q+9aq-6aq^{2}+3q^{2}-24p+48pq-24pq^{2}}{36q^{2}})V_{\alpha}V^{\alpha}\lambda
-pM\lambda=0
\end{eqnarray}
as a quite laborious computation would show. We notice that the conservation laws (\ref{conservationlawspin}-\ref{conservationlawenergy}-\ref{conservationlawtrace}) are in fact verified by the conformal conserved quantities so soon as the conformal matter field equations are satisfied, for both Dirac and ELKO conformal models; putting all together, the whole set of conformal field equations for the conformal Dirac system is such that, because of the complete antisymmetry of the spin, field equations are given by
\begin{eqnarray}
&4k[D_{\rho}P^{\alpha\beta\mu\rho}+V_{\rho}P^{\alpha\beta\mu\rho}
-\frac{1}{2}Q^{\mu}_{\phantom{\mu}\rho\theta}P^{\alpha\beta\rho\theta}
-(\frac{1-q}{3q})(V_{\rho}P^{\rho[\alpha\beta]\mu})]
=\frac{1}{4}\varepsilon^{\mu\alpha\beta\rho}
\overline{\psi}\boldsymbol{\gamma}_{\rho}\boldsymbol{\pi}\psi
\label{Fabbri-spin}\\
\nonumber
&2k[P^{\theta\sigma\rho\alpha}T_{\theta\sigma\rho}^{\phantom{\theta\sigma\rho}\mu}
-\frac{1}{4}g^{\alpha\mu}P^{\theta\sigma\rho\beta}T_{\theta\sigma\rho\beta}
+P^{\mu\sigma\alpha\rho}M_{\sigma\rho}+\\
&+(\frac{1-q}{3q})
(D_{\nu}(2P^{\mu\rho\alpha\nu}V_{\rho})+V_{\nu}(2P^{\mu\rho\alpha\nu}V_{\rho}
-P^{\mu\nu\rho\sigma}Q^{\alpha}_{\phantom{\alpha}\rho\sigma}))]
=\frac{i}{4}(\overline{\psi}\boldsymbol{\gamma}^{\alpha}\boldsymbol{D}^{\mu}\psi
-\boldsymbol{D}^{\mu}\overline{\psi}\boldsymbol{\gamma}^{\alpha}\psi)
\label{Weyl-energy}
\end{eqnarray}
with
\begin{eqnarray}
&i\boldsymbol{\gamma}^{\mu}\boldsymbol{D}_{\mu}\psi
+\frac{i}{2}V_{\mu}\boldsymbol{\gamma}^{\mu}\psi=0
\label{Dirac}
\end{eqnarray}
and they are severely restricted, showing that the complete antisymmetry of the spin contrives not only torsion but also the curvature, while such no constraint is found for ELKO conformal matter field. We notice that constrictions on the curvature may create issues of discontinuity when the torsionless limit is taken \cite{Fabbri:2014kea}, and thus problematic.
%%%%%%%%%%%%%%%%%%%%%%%%%%%%%%%%%%%%%%%%%%%%%%%%%%%%%%%%%%%%%%%%%%%%%%%%%%%%%%%%%%%%%%%%%%%%%%%%%%%
%%%%%%%%%%%%%%%%%%%%%%%%%%%%%%%%%%%%%%%%%%%%%%%%%%%%%%%%%%%%%%%%%%%%%%%%%%%%%%%%%%%%%%%%%%%%%%%%%%%
\section{Effects}
So far, we have exhibited four models through which we could compare ELKO and Dirac fields: 
\begin{enumerate}
 \item \textbf{Standard Background}
\begin{itemize}
 \item \textbf{Dirac Fields}: in this case the torsion-spin field equations are algebraic and explicitly written, so that torsion can be substituted everywhere in terms of the spin leaving spinorial contact interactions; the complete antisymmetry of the spin imposes constraints that are loaded onto torsion which itself becomes completely antisymmetric.
 \item \textbf{ELKO Fields}: in this case the torsion-spin field equations are algebraic but implicitly written, and the problem of the inversion of torsion has not been solved yet; the spin has no particular symmetry and thus torsion has no particular symmetry too.
\end{itemize}
 \item \textbf{Conformal Background}
\begin{itemize}
 \item \textbf{Dirac Fields}: in this case the torsion-spin field equations are differential, and the problem of the inversion of torsion may not even be possible; the complete antisymmetry of the spin imposes constraints that cannot be loaded onto torsion alone and so they are also loaded onto the curvature creating possible discontinuities of the curvature when the torsionless limit is considered.
 \item \textbf{ELKO Fields}: in this case the torsion-spin field equations are differential, and the problem of the inversion of torsion may not even be possible; the spin has no particular symmetry and so also torsion has no particular symmetry and so no constraints are imposed but of course the model may still suffer discontinuity issues.
\end{itemize}
\end{enumerate}
Despite the fact that the Dirac field seems to be naturally conformal we have also seen that when also torsion is considered beside curvature most of this naturalness apparently disappears; albeit ELKO fields may be rendered conformal in torsion-gravity with no such constraints it is nevertheless clear that such a model is astonishingly difficult to investigate: it is as if the mathematical inconsistencies of the Dirac field are quenched for the ELKO field at the price of a practical intractability, in the scheme of conformal torsional-gravitation. Both Dirac and ELKO fields in standard torsion-gravity are tractable: this is due to the fact that the least-order differential character of the theory makes the torsion-spin field equations algebraic in the torsion tensor, and for this reason it is always possible to invert them in order to have torsion explicitly written in terms of torsionless quantities. Still, there are differences.

On the one hand, the complete antisymmetry of the Dirac spin makes it easy to invert the torsion-spin field equations; on the other hand however, the lack of such a symmetry for the ELKO spin forces the inversion to much more complicated: despite the fact that in this case we know that torsion can always be inverted in principle, for the Dirac field such an inversion is immediate while for the ELKO field it is practically difficult and long.

There are shortcuts, however, when considering special symmetries: let us consider for example the cosmological application in which we wish to study FLRW metrics, chosen to be spatially flat as
\begin{eqnarray}
g_{tt}=1,\ \ \ \ \ \ \ \ g_{jj}=-\varsigma^{2}\ \ \ \ j=x,y,z
\end{eqnarray}
where torsion can at most have the time component of the two vector parts $V_{t}$ and $W_{t}$ because any other component will be incompatible with the universal homogeneity and isotropy: in this situation, the ELKO field is given by
\begin{eqnarray}
\begin{tabular}{c}
$\lambda_{+-}=\sqrt{2}\phi\left(\begin{array}{c}
1\\
0\\
0\\ 
-i
\end{array}\!\!\right)$
\end{tabular}\ \ \ \ \mathrm{or} \ \ \ \ 
\begin{tabular}{c}
$\lambda_{-+}=\sqrt{2}\phi\left(\begin{array}{c}
0\\
1\\
i\\ 
0 
\end{array}\!\!\right)$
\end{tabular}
\end{eqnarray}
and if we exploit the presence of free parameters that can be tuned then we may choose $a=-1$ and therefore torsion turns out to have only the time component of the vector part $V_{t}$ and nothing more. In this case it is possible to see that the field equations for the torsion-spin coupling (\ref{torsion-spinELKO}) reduce to be given by the single independent field equation for the torsion-spin trace vector
\begin{eqnarray}
\frac{8}{3}(3A+2B+C-2)V_{\nu}=\partial_{\nu}\lambda^{2}
\label{final}
\end{eqnarray}
which has to be substituted into the field equations for the curvature-energy coupling (\ref{curvature-energyELKO}); when this is done, the gravitational field equations have time-time component that will have only one term containing the combination of parameters $3A+2B+C-2$ at the denominator. If such a combination happens to be very small and negative then the field equations (\ref{curvature-energyELKO}) have time-time component that can be approximated to
\begin{eqnarray}
\frac{\ddot{\varsigma}}{\varsigma}\approx-\frac{1}{(3A+2B+C-2)}\left(\phi\dot{\phi}\right)^{2}
\end{eqnarray}
describing the evolution of the scale factor as a function of the cosmological time, and such that the acceleration $\ddot{\varsigma}/\varsigma$ is positive and large, possibly giving relevant contributions to the accelerated expansion of the late-time universe \cite{FV}.

It is worth remarking that not only in such symmetric situation it becomes easier to invert torsion and have ELKO fields studied in detail, with potentially interesting consequences for cosmological applications, but additionally in these cases ELKO fields might have applications that could escape the reach of the simpler Dirac field: for instance, in the FLRW universe, because of the torsion-spin coupling (\ref{torsion-spinDirac}), the Dirac field is such that $\overline{\psi}\boldsymbol{\gamma}^{\sigma}\boldsymbol{\pi}\psi$ only has the time component; but in then the identity $\overline{\psi}\boldsymbol{\gamma}_{\sigma}\psi
\overline{\psi}\boldsymbol{\gamma}^{\sigma}\boldsymbol{\pi}\psi=0$ tells that either $\overline{\psi}\boldsymbol{\gamma}^{0}\boldsymbol{\pi}\psi=0$ or $\overline{\psi}\boldsymbol{\gamma}_{0}\psi=0$ and thus respectively, either also the temporal component of $\overline{\psi}\boldsymbol{\gamma}^{\sigma}\boldsymbol{\pi}\psi$ vanishes or $\psi^{\dagger}\psi$ vanishes, and again either spin is equal to zero or the fermion itself is equal to zero. In any case, the spin would vanish and with it, torsion would vanish too, rendering the effects of torsion at cosmological scales not only irrelevant, but exactly null; in this sense, there can be no Dirac field with torsion in cosmology. Surprisingly, this fact is constantly overlooked throughout the present literature.

The differences between ELKO and Dirac fields lie on the fact that ELKO is a higher-order derivative field while Dirac is a least-order derivative field, that is the ELKO spin is differential while the Dirac spin is algebraic: so whichever constraint a given symmetry places on torsion, the torsion-spin field equations (\ref{final}) load it onto the derivatives of the field while the torsion-spin field equations (\ref{torsion-spinDirac}) load it onto the field itself, so that limitations are imposed in the former case for the dynamics of the field while in the latter case for the field itself. Once again, it seems that the mathematical incompatibility between some symmetry and the Dirac field are quenched for the ELKO field with those symmetries at the price of practical difficulties, but at least in standard torsion-gravity these difficulties can be overcome and the higher manoeuvrability of ELKO fields compared to Dirac fields is a clear richness.

As a finally remark, it is important to stress that such a comparison between ELKO and Dirac fields was essentially based on the presence of the torsion-spin field equations, and therefore no such differences would have ever been apparent it torsion were to be arbitrarily abolished from the very foundations of the geometry.
%%%%%%%%%%%%%%%%%%%%%%%%%%%%%%%%%%%%%%%%%%%%%%%%%%%%%%%%%%%%%%%%%%%%%%%%%%%%%%%%%%%%%%%%%%%%%%%%%%%
%%%%%%%%%%%%%%%%%%%%%%%%%%%%%%%%%%%%%%%%%%%%%%%%%%%%%%%%%%%%%%%%%%%%%%%%%%%%%%%%%%%%%%%%%%%%%%%%%%%
\section*{Conclusion}
In this paper, we have considered the standard and the conformal approaches to the theory of torsion-gravitation in order to draw a comparison between ELKO and Dirac fields based on the analysis of their torsion-spin coupling field equations and their properties: we have seen that in conformal gravity, torsion is linked to the spin differentially, that is torsion is propagating, while in standard gravity, the torsion-spin field coupling is algebraic, that is torsion does not propagate, and that while for ELKO fields the spin has a derivative structure for the Dirac field the spin is only algebraic: thus, we have had the opportunity to appreciate the intrinsic complexity of the propagating torsion of the conformal approach compared to the non-propagating torsion of the standard approach, and in both cases it has been possible to witness that whether in terms of inconsistencies in the mathematical structure of incompatibility with a given underlying symmetry, the Dirac field is much less versatile than the ELKO field; that is, torsion has the tendency to constrain much more easily the Dirac field than the ELKO field. In the perspective of torsion, the ELKO field is therefore characterized by a more flexible dynamics compared to the Dirac field, and this constitutes a certain advantage whenever it can be investigated; the disadvantage is of course the fact that in general it is more difficult to investigate ELKO than Dirac fields. And finally, we have drawn a concluding remark stressing that none of these differences would have been visible if we decided to have torsion neglected at the beginning.
%%%%%%%%%%%%%%%%%%%%%%%%%%%%%%%%%%%%%%%%%%%%%%%%%%%%%%%%%%%%%%%%%%%%%%%%%%%%%%%%%%%%%%%%%%%%%%%%%%%

\

\noindent\textbf{Acknowledgments.} The authors would like to thank Prof.~D.~V.~Ahluwalia, Dr.~R.~da Rocha, Dr.~C.~Y.~Lee, Dr.~J.~M.~Hoff da Silva for their kind invitation to write this paper for International Journal of Modern Physics D, Special Issue: ELKO and Mass-Dimension-One Fermions, as Proceedings of the Second Workshop on ELKO, held in Brazil, in May 2014; one of the authors, L.~F., is thankful to the University of Genoa, Italy, for financial support.
%%%%%%%%%%%%%%%%%%%%%%%%%%%%%%%%%%%%%%%%%%%%%%%%%%%%%%%%%%%%%%%%%%%%%%%%%%%%%%%%%%%%%%%%%%%%%%%%%%%
%%%%%%%%%%%%%%%%%%%%%%%%%%%%%%%%%%%%%%%%%%%%%%%%%%%%%%%%%%%%%%%%%%%%%%%%%%%%%%%%%%%%%%%%%%%%%%%%%%%

%%%%%%%%%%%%%%%%%%%%%%%%%%%%%%%%%%%%%%%%%%%%%%%%%%%%%%%%%%%%%%%%%%%%%%%%%%%%%%%%%%%%%%%%%%%%%%%%%%%
\end{document}